\documentclass[%
 reprint,
 amsmath,amssymb,
 aps,
prl,
floatfix,
]{revtex4-1}
\usepackage{epsf}
\usepackage{graphicx}
\usepackage{dcolumn}
\usepackage{bm}

\begin{document}


\title{Experimental evidence for non-axisymmetric magnetorotational instability
in a rotating liquid metal exposed to 
an azimuthal magnetic field}

\author{Martin Seilmayer}
\author{Vladimir Galindo}
\author{Gunter Gerbeth}
\author{Thomas Gundrum}
\author{Frank Stefani}%
\email{F.Stefani@hzdr.de}
\affiliation{Helmholtz-Zentrum Dresden-Rossendorf, P.O. Box 510119, D-01314 Dresden, Germany}%

\author{Marcus Gellert}
\author{G\"unther R\"udiger}
\author{Manfred Schultz}
\affiliation{Leibniz-Institut f\"ur Astrophysik Potsdam, 
An der Sternwarte 16, D-14482 Potsdam, Germany}%

\author{Rainer Hollerbach}
\affiliation{Department of Applied Mathematics, University of Leeds, 
Leeds LS2 9JT, U.K.}%

\date{\today}

\begin{abstract}
The azimuthal version of the magnetorotational instability (MRI) is a
non-axisymmetric instability of a hydrodynamically stable differentially
rotating flow under the influence of a purely or predominantly azimuthal
magnetic field. It may be of considerable importance for destabilizing
accretion disks, and plays a central role in the concept of the MRI dynamo.
We report the results of a liquid metal Taylor-Couette experiment that
shows the occurrence of an azimuthal MRI in the expected range of Hartmann
numbers.
\end{abstract}

\pacs{47.20.Ft, 47.65.-d}
\maketitle


The magnetorotational instability (MRI) is widely accepted as the main source
of turbulence and angular momentum transport in accretion disks around
protostars and black holes. Although discovered by Velikhov \cite{VELI59} as
early as 1959, its relevance for the evolution of stellar systems, X-ray
binaries, and active galactic nuclei was only recognized by Balbus and Hawley
in 1991 \cite{BH91}. While most of the early MRI studies had considered a
uniform axial magnetic field threading the flow (non-zero net-flux), the
recent focus \cite{MNRAS} has shifted slightly to the case of azimuthal fields (zero
net-flux). One reason for this lies in the interesting concept of a
subcritical MRI dynamo, in which the MRI-triggering field is partly sustained
by the MRI-driven turbulence itself \cite{BRRIHE}. 
Another possible application is related to the high 
values of ''artificial viscosity'' that are needed to 
explain the slowing down of stellar cores after the collapse 
towards their red giant stage \cite{EGG}.

In magnetohydrodynamic stability problems of this kind, the magnetic Prandtl
number $\rm Pm$, the ratio of the fluid's kinematic viscosity to its magnetic
diffusivity, can play a crucial role. For $\rm Pm \ge 1$ both non-zero and zero
net-flux versions of the MRI operate very effectively and robustly, whereas for
$\rm Pm \ll 1$ both are far more delicate, involving not only numerical
convergence issues, but also real physical effects such as the role of
stratification or boundary conditions \cite{FRSTRKAP}.

\begin{figure}[t]
\begin{center}
\epsfxsize=7.0cm\epsfbox{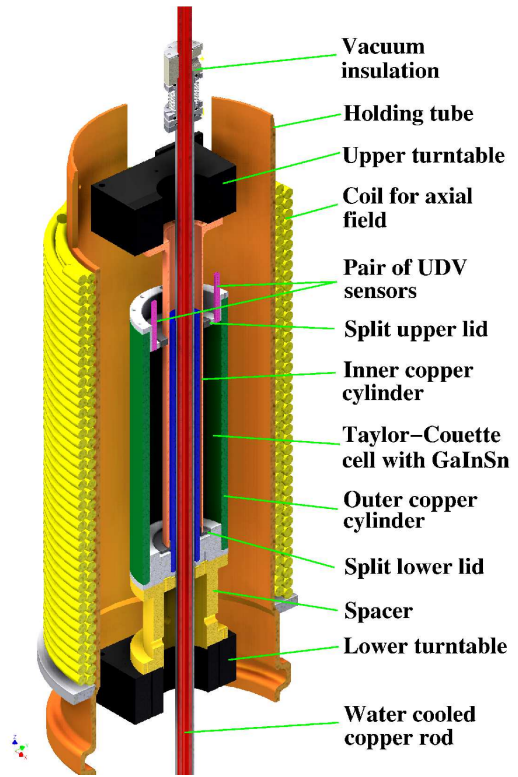}
\vspace{2mm}
\caption{Sketch of the experimental set-up.}
\end{center}
\end{figure}

The discovery of the helical MRI (HMRI) by Hollerbach and R\"udiger in 2005
\cite{HR05} spurred additional interest in the small $\rm Pm$ limit. For an
appropriate combination of axial and azimuthal magnetic fields, the HMRI was
shown to work even in the inductionless limit $\rm Pm \rightarrow 0$, since it
depends only on the Reynolds number $\rm Re$ and the Hartmann number $\rm Ha$.
This is in contrast with the standard MRI (SMRI) which requires both the
magnetic Reynolds number $\rm Rm= \rm Pm \rm Re$ and the Lundquist number
$\rm S= {\rm Pm}^{1/2} \rm Ha$ to be at least $O(1)$, and is correspondingly
difficult to observe in the laboratory \cite{DANHAN}.

It is of interest also to consider the possibility of a (non-axisymmetric) MRI
operating in a purely azimuthal field \cite{OGILVIE}, a configuration that has
come to be known as the azimuthal MRI (AMRI) \cite{AMRI1}. It was initially
believed that the AMRI operates only in the same $\rm(Rm,S)>O(1)$ parameter
regime as the SMRI, and would be experimentally unobtainable. However, in 2010
it was discovered that for sufficiently steep rotation profiles the AMRI
switches to the same inductionless $\rm(Re,Ha)$ parameter values as the HMRI
\cite{TEELUCK}. It is this inductionless version of the AMRI that will be
explored in this Letter.

The question of which parameters, $\rm(Rm,S)$ or $\rm(Re,Ha)$, are the relevant
ones, and how that might vary depending on the steepness of the rotation profile,
is also of potential astrophysical significance, since the Keplerian profile
$\Omega(r)\propto r^{-3/2}$ that is of greatest interest in accretion disks is
considerably shallower than the limiting Rayleigh value $\Omega(r)\propto r^{-2}$.
For the azimuthal field profile $B_{\phi} \propto r^{-1}$ both the HMRI as well
as the AMRI have switched from $\rm(Re,Ha)$ back to $\rm(Rm,S)$ for rotation
profiles as shallow as Keplerian, as was first noted for the HMRI by Liu et al.
\cite{LIU} and generalized to 
higher $m$ modes by Kirillov et al. \cite{KSF12}. 
However, if the 
field profiles are taken only slightly shallower than
$B_{\phi} \propto r^{-1}$, both the HMRI and the AMRI have recently been shown
\cite{KS13} to scale with $\rm(Re,Ha)$ even for Keplerian rotation
profiles. The astrophysical importance of these $\rm(Rm,S)$ versus $\rm(Re,Ha)$
scaling laws thus continues to be an open question.

\begin{figure}[t]
\begin{center}
\epsfxsize=8.0cm\epsfbox{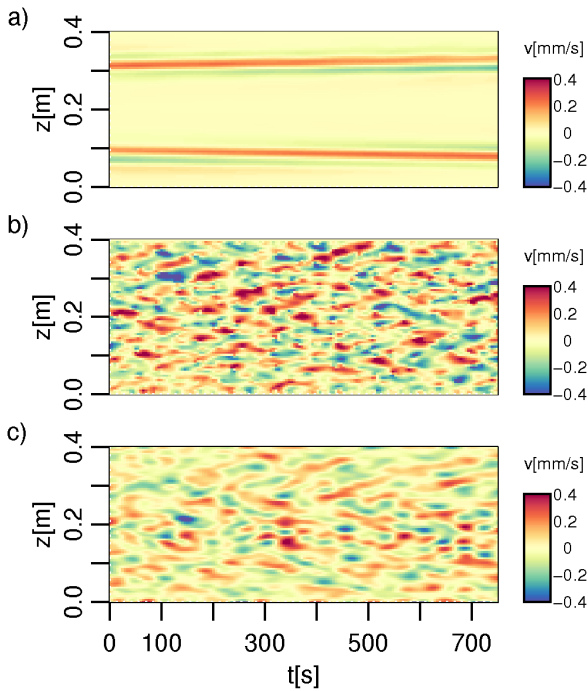}
\vspace{2mm}
\caption{Velocity perturbation $v_z(m=1,z,t)$ for $\mu=0.26$, $\rm Re=1480$,
and ${\rm Ha}=77$. (a) 3D simulation for ideal axisymmetric field. (b)  3D simulation
for realistic field. (c) Experimental results.}
\end{center}
\end{figure}

\begin{figure}[t]
\begin{center}
\epsfxsize=8.0cm\epsfbox{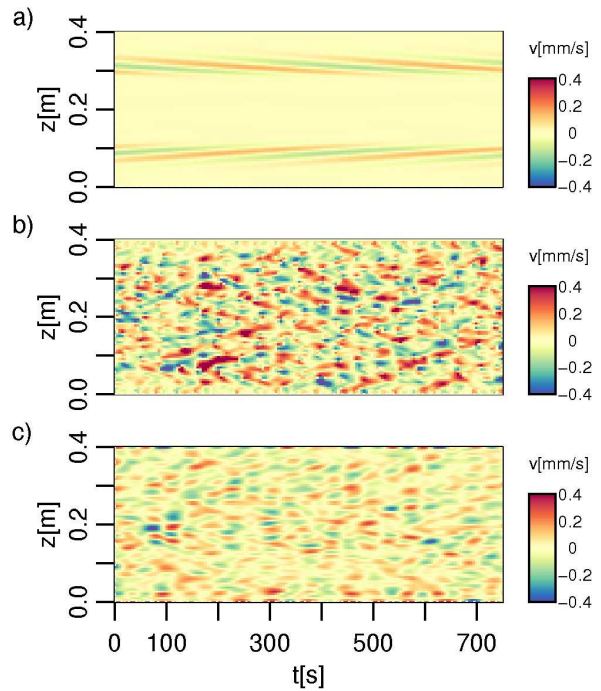}
\vspace{2mm}
\caption{As in Fig.\ 2, but for ${\rm Ha}=110$.}
\end{center}
\end{figure}

In order to study this non-axisymmetric AMRI, we utilize a slightly modified
version of the PROMISE facility which has previously been used for investigations
of the HMRI \cite{PRLNJPPRE}. The main part of PROMISE is a cylindrical vessel
(Fig. 1) made of two concentric copper cylinders enclosing a cylindrical
volume of width $d=r_{\rm out} - r_{\rm in}=$ 40 mm, between the radii
$r_{\rm in}=40$ mm and $r_{\rm out}=80$ mm, and a height of 400 mm. This
cylindrical volume is filled with the liquid eutectic alloy
Ga$^{67}$In$^{20.5}$Sn$^{12.5}$ for which 
${\rm Pm}=1.4 \times 10^{-6}$. Both 
the upper and lower end-caps of the cylindrical
volume are formed by two plastic rings, separated at $r=56$ mm, the inner and
outer ring rotating with the inner and outer cylinders, respectively. 

\begin{figure}[h]
\begin{center}
\epsfxsize=8.0cm\epsfbox{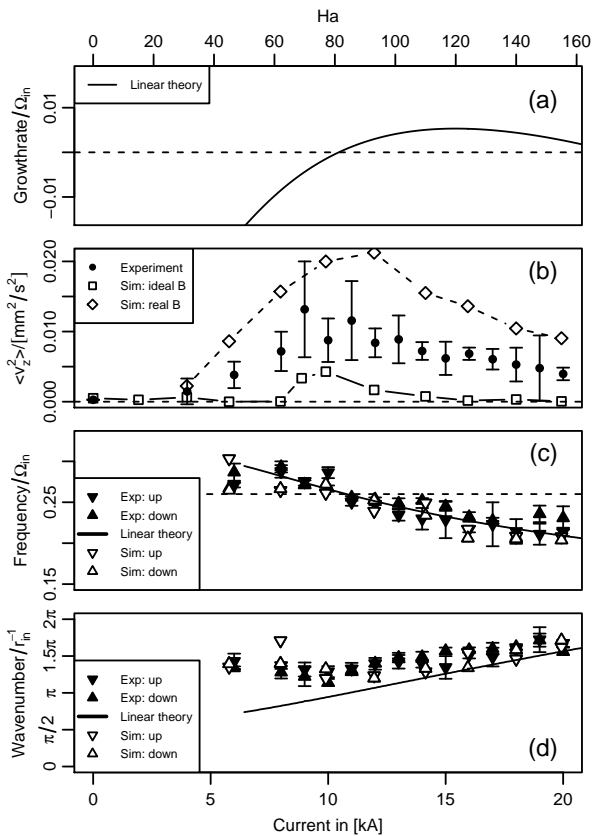}
\vspace{2mm}
\caption{Results for ${\rm Re}=1480$. (a) 
Growth rate from a 1D linear stability code, (b) mean squared 
velocity perturbation from experiment and two different 
3D simulations for idealized and real magnetic field geometry, (c) 
angular drift frequency and (d) wavenumber from experiment, 
from the 1D linear stability code, 
and from the 3D simulation with real field geometry.
The error bar of the velocity perturbation corresponds to
an 85 percent confidence level. 
The ``up'' and ``down'' values in (c) and (d), which refer
to the travel direction of the velocity perturbations as exemplified 
in  Figs. 2 and 3,
are determined by a center-of-gravity method applied to the 
2D-FFT of the data.}
\end{center}
\end{figure}

\begin{figure}[h]
\begin{center}
\epsfxsize=8.0cm\epsfbox{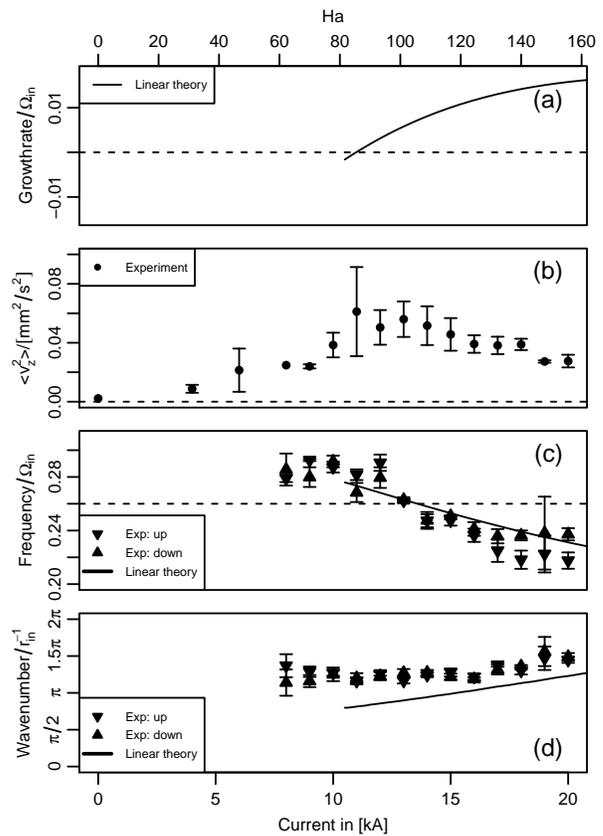}
\vspace{2mm}
\caption{As in Fig.\ 4 (except the 3D simulation data),
but for ${\rm Re}=2960$.}
\end{center}
\end{figure}

The magnetic field configuration is basically identical to that of the
previous PROMISE experiments, apart from a significant enhancement of the
power supply that now allows for currents in the central copper rod of up to
20 kA. This value is approximately double the expected critical value for
the onset of AMRI \cite{TEELUCK,MNRAS}. The central rod can become quite hot,
and was therefore thermally insulated by a vacuum tube to prevent any
convection effects in the fluid. This vertical copper rod is connected to the
power source by two horizontal rods at a height of 0.8 m below the bottom and
above the top of the cylindrical volume. The slight deviation from a purely
axisymmetric $B_{\phi}(r)$ that arises from this asymmetric wiring will play
an important role below. We further mention that the coil for the production
of the axial field $B_z$ was left in place, although it was not used for the
particular experiments reported in this Letter.
In any case, there is 
no electrical current applied to the liquid metal, in contrast to 
previous experiments on the pinch-type Tayler instability 
\cite{SEILMAYER}.

With $B_z$ being set to zero, the AMRI is completely governed by only three
non-dimensional parameters, the Reynolds number 
${\rm Re}:=\Omega_{\rm in} d r_{\rm in}/\nu$,
the ratio of outer to inner angular frequencies
$\mu:=\Omega_{\rm out}/\Omega_{\rm in}$,
and the Hartmann number characterizing the azimuthal magnetic field:
${\rm Ha}:=B_{\phi}(r_{\rm in}) (r_{\rm in} d \sigma/\rho \nu)^{1/2}$.
For converting between dimensional and non-dimensional quantities we can use
the following relations:
${\rm Re}=4710 \; \Omega_{\rm in}/s^{-1}$ and
${\rm Ha}=7.77 \; I_{\rm rod}/{\rm kA}$.

The measuring instrumentation consists of two Ultrasonic Doppler Velocimetry
(UDV) transducers (from Signal Processing SA) working at a frequency around
3.5 MHz, which are fixed into the outer plastic ring, 12 mm away from the outer
copper wall, and flush mounted at the interface to the GaInSn. The signals from
these sensors are transferred from the rotating frame of the outer cylinder to
the laboratory frame by means of a slip ring contact which is situated below
the vessel (not shown in Fig.\ 1). The UDV provides profiles of the axial
velocity $v_z$ along the beam-lines parallel to the axis of rotation. The
spatial resolution in the axial direction is around 1 mm; the time resolution
is 2 sec.

From previous work \cite{TEELUCK,MNRAS} we anticipate that the AMRI starts 
at ${\rm Ha}\simeq 80$  and manifests
itself as a non-axisymmetric ($m=\pm 1$) spiral velocity structure that rotates
around the vertical axis with an angular frequency close to that of the
outer cylinder. This $m=\pm 1$ mode can be identified by taking the difference
of the signals of the two UDV transducers, although the observed frequency in
the co-rotating frame of the sensors will be rather small. Among other numerical
tools \cite{MNRAS}, we have used the OpenFoam library, enhanced by a Poisson
solver for the determination of the induced electric potential (see \cite{WEBER}
for details), in order to simulate the AMRI for the true geometry of the
facility and the real $\rm Pm$ of GaInSn. The velocity structure simulated in
this way can then be transformed to the co-rotating frame in order to compare
the resulting velocity pattern with the experimentally observed one.

This is done in Figs.\ 2 and 3, for $\mu=0.26$, $\rm Re=1480$, and ${\rm Ha}=77$
and ${\rm Ha}=110$, respectively. For the idealized case of a perfectly
axisymmetric $B_{\phi}(r)$, Fig.\ 2a illustrates the simulated $m=\pm 1$
projection of the axial velocity perturbation $v_z(z,t)$ in dependence on time
$t$ and vertical position $z$, when virtually transformed to the co-rotating
frame of the UDV sensors. The resulting ``butterfly'' pattern represents a
spiral, rotating slightly faster than the outer cylinder, whose energy is
concentrated approximately in the middle parts of the upper and lower halves of
the cylinder. In Fig.\ 2b we show the simulation for the real geometry of the
applied magnetic field, including its slight symmetry-breaking due to the
one-sided wiring (the deviation is about 5 percent at the inner radius, and 10
percent at the outer radius). The effect is remarkable: the formerly clearly
separated spiral structures now also fill the middle part of the cylinder and
penetrate into the other halves  (a comparable effect in which
individual left- and right spiral waves are replaced by 
interpenetrating spirals has been investigated 
in connection with the double Hopf bifurcation in a 
corotating spiral Poiseuille flow \cite{AVILA}.) 
The corresponding velocity pattern observed in
the experiment is shown in Fig.\ 2c; the similarity to the realistic simulation
in Fig.\ 2b is striking. Note that in both Figs.\ 2b and 2c we have filtered out
those components of the $m=\pm 1$ modes that are stationary in the laboratory
frame, since they are a direct consequence of the external symmetry breaking,
without any relation to the AMRI mode as such.

The same procedure is documented for ${\rm Ha}=110$ in Fig. 3. Again, Fig.\ 3a
shows the numerically computed pattern for the perfectly axisymmetric case. The
``butterfly diagram'' has now changed its direction, meaning that the spiral
rotates slightly slower than the outer cylinder. The more realistic simulation
in Fig.\ 3b shows the interpenetration of the spirals of the upper and lower
halves of the cylinder, which is also qualitatively confirmed by the
experimental data in Fig.\ 3c.

By analyzing a total of 102 experimental runs similar to those documented in
Figs.\ 2c and 3c, we have extracted the dependence of various quantities on
$\rm Ha$. For $\mu=0.26$ and $\rm Re=1480$, Fig.\ 4a shows the theoretical
growth rate of the AMRI as determined by a 1D-eigenvalue solver for the
infinite length system \cite{MNRAS}. In Fig.\ 4b we show then the squared rms
of the UDV-measured velocity perturbation $v_z(m=1,z,t)$ and compare them with
the numerically determined ones for the idealized axisymmetric and the realistic
applied magnetic fields. Whereas the growth rate in Fig.\ 4a and the numerical
rms results under the axisymmetric field condition give a consistent picture
with a sharp onset of AMRI at $\rm Ha\simeq 80$ \cite{MNRAS,TEELUCK}, the slight 
symmetry breaking of
the field leads, first, to some smearing out of the rms for lower $\rm Ha$ and,
second, to a significant increase of the rms velocity value, with a  reasonable
correspondence of numerical and experimental values. The remaining 
deviation of the rms value might have to do with the 
smoothing and filtering processes that are necessary due to the high noise 
level of the raw data from the UDV (which is indeed at the edge of 
applicability here), as well as with some compromises made in the 
numerical simulation, in particular with respect to the complicated 
electrical boundary conditions.

The dependence of the numerically and experimentally determined normalized drift
frequency on $\rm Ha$ is shown in Fig.\ 4c. AMRI represents an $m=\pm 1$ spiral
pattern that rotates approximately with the rotation rate of the outer cylinder
\cite{MNRAS}. There is still some deviation from perfect co-rotation, with a
slightly enhanced frequency for lower $\rm Ha$ and a slightly reduced frequency
for higher $\rm Ha$, which can be identified both in the linear theory and in
the experimental data. The corresponding wavenumbers are 
presented in Fig.\ 4d.

For a doubled rotation rate, i.e. $\rm Re=2960$, Fig.\ 5 shows the same
quantities as in Fig.\ 4, except with the numerical predictions restricted to
those of the 1D eigenvalue solver, since 3D simulations already become extremely
expensive in this case. Still, the qualitative behaviour of the rms and the
frequency agrees well with that at the lower rotation rates (Fig.\ 4).

In summary, we have shown that AMRI occurs in a hydrodynamically stable
differential rotational flow of a liquid metal when it is exposed to a
dominantly azimuthal magnetic field. The critical Hartmann number for the
onset of AMRI is close to the numerically predicted value of approximately 80.
The dependence of the rms, the drift frequency, and the axial wavenumber 
of the non-axisymmetric velocity
perturbations on $\rm Ha$ turned out to be in good agreement with numerical
predictions, especially if the latter incorporate the surprisingly strong
effect of the slight symmetry-breaking of the externally applied magnetic 
field. This underlines the importance of 3D codes, working at 
realistic $\rm Pm$, for a detailed understanding of experimental results.
Presently, experimental and numerical work is going on to scrutinize the 
dependence of the AMRI on the ratio $\mu$ of outer to inner cylinder 
rotation rates. The main focus here is on
whether the (modified) AMRI could possibly extend to rotation profiles as 
flat as the Keplerian one.

Significantly more experimental effort is needed to study the 
influence of an additionally
applied $B_z$, which breaks the symmetry between the $m=1$ and $m=-1$ modes
\cite{TEELUCK}. When increasing $B_z$ even 
further (at $B_z \simeq 0.05 B_{\phi}(r_{\rm in})$, see Fig. 3 of \cite{TEELUCK}), 
we should also be able to
observe the transition from the $m=\pm 1$ AMRI mode back to the previous $m=0$
HMRI mode which can be identified in the sum of the signals
of the two UDV transducers \cite{PRLNJPPRE}. A more ambitious project, 
planned within the framework of the
DRESDYN project \cite{DRESDYN}, will comprise a large liquid sodium experiment
for the combined investigation of SMRI, HMRI, AMRI, and the current-driven
Tayler instability \cite{SEILMAYER}.

This work was supported by the Helmholtz-Gemeinschaft Deutscher
Forschungszentren (HGF) in the framework of the Helmholtz Alliance LIMTECH, as
well as by the Deutsche Forschungsgemeinschaft under SPP 1488 (PlanetMag). M.S.
would like to thank Hans Georg Krauth\"auser for many discussions on data
analysis problems. Technical support by Steffen Borchardt, Heiko Kunath, and
Bernd Nowak is gratefully acknowledged.

\end{document}